\documentstyle[12pt,aps,amsfonts,epsf]{revtex}

\def\bm#1{\mbox{\boldmath $#1$}}

\newcommand{\ket}[1]{\mbox{$|{#1}\rangle$}}

\def\Tr{{\rm Tr}}
\begin{document}

\title{
Eigenvalues, Peres' separability condition and entanglement 
\thanks{Supported by the National Natural Science Foundation of China under Grant No. 69773052.}}
\author{An Min WANG$^{1,2,3}$}
\address{CCAST(World Laboratory) P.O.Box 8730, Beijing 100080$^1$\\
and Laboratory of Quantum Communication and Quantum Computing\\
University of Science and Technology of China$^2$\\
Department of Modern Physics, University of Science and Technology of China\\
P.O. Box 4, Hefei 230027, People's Republic of China$^3$}
\maketitle
\centerline{(\it Revised Version)}

\bigskip

\begin{abstract}
The general expression with the physical significance and positive definite condition of the eigenvalues of $4\times 4$ Hermitian and trace-one matrix are obtained. This implies that the eigenvalue problem of the $4\times 4$ density matrix is generally solved. The obvious expression of Peres' separability condition for an arbitrary state of two qubits is then given out and it is very easy to use. Furthermore, we discuss some applications to the calculation of the entanglement, the upper bound of the entanglement, and a model of the transfer of entanglement in a qubit chain through a noisy channel.



\end{abstract}
\bigskip


The density matrix (DM) was introduced by J. von Neumann to describe the statistical concepts in quantum mechanics \cite{Neumann}. The main virtue of DM is its analytical power in the construction of the general formulas and in the proof of the general theorems. The evaluation of averages and probabilities of the physical quantities characterizing a given system is extremely cumbersome without the use of density matrix techniques. Recently, the application of DM has been gaining more and more importance in the many fields of physics. For example, in the quantum information and quantum computing\cite{QC}, DM techniques have become an important tool for describing and characterizing the measure of entanglement, purification of entanglement and encoding \cite{Bennett1,Deutsch,Horodecki,Wootters}. However, even if DM is a simple enough $4\times4$ dimensional one, to write a general expression of its eigenvalues in a compact form with the physical significance seems not to be a trivial problem. Although one has known the theory of quartic equation, this is still difficult since DM has 15 independent parameters. Actually, we need a physical closed form for it but not a mathematical closed form only with formal meaning. This letter is just devoted to this fundamental problem in quantum mechanics. It successfully finds out the general expression of the eigenvalues of a $4\times 4$ density matrix with a clear physical significance and in a compact form. Thus, an obvious expression of Peres' separability condition is derived clearly. This provides a very easy and direct way to use it. Moreover, some important applications to the entanglement and separability in the quantum information, such as the calculation of the entanglement, the upper bound of the entanglement and the transfer of the entanglement, are discussed constructively. 

The elementary unit of quantum information is so-called ``qubit" \cite{Schumacher}. A single qubit can be envisaged as a two-state quantum system such as a spin-half or a two-level atom. A pair of qubits forms the simplest quantum register which can be expressed by a $4\times 4$ density matrix. Just is well known, the eigenvalues of DM of two qubits are closely related with its entanglement and separability. For example, Wootters gave a measure of the entanglement in terms of the eigenvalues \cite{Wootters}, and Peres' separability condition depends on the positive definite property of the partial transpose of DM. Therefore, it is very interesting and essentially important to know what is the general expression of eigenvalues of DM of two qubits in an arbitrary state. 

DM of two qubits can be written as
\begin{equation}
\rho=\frac{1}{4}\sum_{\mu,\nu=0}^3 a_{\mu\nu}\sigma_\mu\otimes\sigma_\nu,\label{Rhoe}
\end{equation}
where $\sigma_0$ is two dimensional identity matrix and $\sigma_i$ is the usual Pauli matrix. $\rho=\rho^\dagger$ (Hermitian) leads to $a_{\mu\nu}$ be the real numbers,  ${\rm Tr}\rho=1$ (trace-one) requires $a_{00}=1$, and from the eigenvalue of Pauli matrix it follows that $-1\leq a_{\mu\nu}\leq 1$. Moreover, it is easy to get
\begin{equation}
a_{\mu\nu}={\rm Tr}(\rho\sigma_\mu\otimes\sigma_\nu).\label{CA}
\end{equation}
Note that Eq.(\ref{Rhoe}) does not involve with the positive definite condition for $\rho$. In order to find out the general expression of the eigenvalues of DM, we first give out the following two lemmas. 
\medskip

\noindent {\it Lemma One}\ The form of the characteristic polynomial of a $4\times 4$ Hermitian and trace-one matrix $\Omega$ is
\begin{equation}
b_0+b_1\lambda + b_2\lambda^2-\lambda^3+\lambda^4,
\end{equation}
where the coefficients $b_0,b_1$ and $b_2$ are defined by
\begin{eqnarray}
b_0&=&\frac{1}{64}[1-\bm{\xi}_A^2 \bm{\xi}_B^2-(A\bm{\xi}_A)^2-(A\bm{\xi}_B)^2 \nonumber \\
& &+2\bm{\xi}_A^T A \bm{\xi}_B +(({\rm Tr}A)^2-{\rm Tr}A^2)\bm{\xi}_A \cdot\bm{\xi}_B \nonumber \\
& &+2\bm{\xi}_B^T A^2\bm{\xi}_A-2{\rm Tr}A\;\bm{\xi}_B^T A\bm{\xi}_A-(\bm{a}_1\times\bm{a}_2)^2 \nonumber\\
& &-(\bm{a}_2\times\bm{a}_3)^2-(\bm{a}_3\times\bm{a}_1)^2-2(\bm{a}_1\times\bm{a}_2)\cdot\bm{a}_3] \nonumber\\
& &-\frac{1}{16}[{\rm Tr}\Omega^2-({\rm Tr}\Omega^2)^2],\\
b_1&=&\frac{1}{8}[2{\rm Tr}\Omega^2-1-\bm{\xi}_A^T A \bm{\xi}_B+(\bm{a}_1\times\bm{a}_2)\cdot\bm{a}_3],\\
b_2&=&\frac{1}{2}(1-{\rm Tr}\Omega^2).
\end{eqnarray}
In the above equations, we have introduced the polarized vectors of the reduced density matrices $\bm{\xi}_A=(a_{10},a_{20},a_{30})$, $
\bm{\xi}_B=(a_{01},a_{02},a_{03})$; the space Bloch's vector $\bm{a}_1=(a_{11},a_{12},a_{13})$, $\bm{a}_2=(a_{21},a_{22},a_{23})$, $ \bm{a}_3=(a_{31},a_{32},a_{33})$; 
 and the polarized rotation matrix $A=\{a_{ij}\}\; (i,j=1,2,3)$.
Note that $\bm{\xi}_{\{A,B\}}$ is viewed as a column vector and its transpose $\bm{\xi}^{\rm T}_{\{A,B\}}$ is then a row vector. The physical meaning of $3\times 3$ matrix $A$ can be seen in my paper \cite{My0}. Again, the positive definite condition has not been used here and $a_{\mu\nu}$ is defined just as Eq.(\ref{CA}).

\medskip
\noindent{\it Lemma Two}\ If a $4\times 4$ Hermitian and trace-one matrix $\Omega$ has $m$ non-zero eigenvalues, then
\begin{equation}
{\rm Tr}\Omega^2\geq \frac{1}{m}.
\end{equation}
If $\Omega$ is positive definite, then
\begin{equation}
{\rm Tr}\Omega^2\leq 1.
\end{equation}

To prove Lemma one, we need to use the physical ideas to arrange those coefficients from the characteristic determinant into a compact form. And, it leads to the general expression of the eigenvalues of DM with the physical significance for their applications. Lemma two can be obtained by the standard method to find the extremum. Based on the theory of the quartic equation, we have

\medskip
\noindent{\it Theorem One}\ The eigenvalues of the $4\times 4$ Hermitian and trace-one matrix $\Omega$ are
\begin{eqnarray}
\lambda^{\pm}(-)&=&\frac{1}{4}-\frac{1}{4\sqrt{3}}(4{\rm Tr}\Omega^2-1+8c_1\cos\phi)^{1/2} \nonumber\\
& &\pm\frac{1}{2\sqrt{6}} \left[4{\rm Tr}\Omega^2-1-4c_1\cos\phi+\frac{3\sqrt{3}(1+8b_1-2{\rm Tr}\Omega^2)}{\sqrt{4{\rm Tr}\Omega^2-1+8c_1\cos\phi}}\right]^{1/2},\\
\lambda^\pm(+)&=&\frac{1}{4}+\frac{1}{4\sqrt{3}}(4{\rm Tr}\Omega^2-1+8c_1\cos\phi)^{1/2}\nonumber \\
& &\pm\frac{1}{2\sqrt{6}}\left[4{\rm Tr}\Omega^2-1-4c_1\cos\phi-\frac{3\sqrt{3}(1+8b_1-2{\rm Tr}\Omega^2)}{\sqrt{4{\rm Tr}\Omega^2-1+8c_1\cos\phi}}\right]^{1/2},
\end{eqnarray}
where
\begin{eqnarray}
\cos3\phi&=&\frac{c_2}{2c_1^3}=4\cos^3\phi-3\cos\phi\label{3Phi}, \\
c_1&=&\sqrt{12b_0+3b_1+b_2^2},\\
c_2&=&27b_1^2+b_0(27-72b_2)+9b_1b_2+2b_2^3.
\end{eqnarray}
In the above equations, we have assumed $c_1\neq 0, c_2\neq 0$. Note that $c_2^2-4c_1^6=-27(\lambda_1-\lambda_2)^2(\lambda_1-\lambda_3)^2(\lambda_1-\lambda_4)^2(\lambda_2-\lambda_3)^2(\lambda_2-\lambda_4)^2(\lambda_3-\lambda_4)^2$ is non-positive. Thus, $4c_1^6\geq c_2^2\geq 0$, and so $c_1$ is real. If there is any repeated root ($c_2^2-4c_1^6=0$), $\phi=0$ or $\pi/3$ since $c_2=\pm 2c_1^3$. In fact, Eq.(\ref{3Phi}) implies that
\begin{eqnarray}
\cos\phi&=&\frac{c_1}{2^{2/3}(c_2+\sqrt{c_2^2-4c_1^6})^{1/3}}+\frac{(c_2+\sqrt{c_2^2-4c_1^6})^{1/3}}{2\times 2^{1/3}c_1},\\
\phi&=&{\rm Arg}\left[\left(c_2+\sqrt{c_2^2-4c_1^6}\right)^{1/3}\right]
\end{eqnarray}
Obviously, if $c_2<0$, then $\pi/6<\phi\leq \pi/3$, and if $c_2>0$, then $0\leq\phi<\pi/6$. 

Now consider the case that $c_1$ or/and $c_2$ are equal to zero. We discuss it by two steps. First, suppose $b_2=3/8$ or ${\rm Tr}\Omega^2=1/4$. If only $c_1=0$ or $c_2=0$, then some of these eigenvalues will be complex numbers. This is contradict with Hermitian property of DM. So this conclusion means that both $c_1$ and $c_2$ have to be zero together. Thus, we can obtain that all the eigenvalues are $1/4$. Second, suppose $b_2\neq 3/8$ or ${\rm Tr}\Omega^2\neq 1/4$. We have to analysis the possibilities stated as the following.  

For only $c_1=0$ or $12b_0+3b_1+b_2^2=0$, again from $c_2^2=c_2^2-4c_1^6=-27(\lambda_1-\lambda_2)^2(\lambda_1-\lambda_3)^2(\lambda_1-\lambda_4)^2(\lambda_2-\lambda_3)^2(\lambda_2-\lambda_4)^2(\lambda_3-\lambda_4)^2\leq 0$, we have that $c_2$ has to be zero since it is real. So only $c_1=0$ is impossible. 

For only $c_2=0$ or $27b_1^2+b_0(27-72b_2)+9b_1b_2+2b_2^3=0$, the eigenvalues of the $4\times 4$ Hermitian and trace-one matrix are then
\begin{eqnarray}
\lambda^\pm(-)&=&\frac{1}{4}-\frac{1}{4\sqrt{3}}\sqrt{4{\rm Tr}\Omega^2-1}\pm\frac{1}{2\sqrt{6}}\left(4{\rm Tr}\Omega^2-1+\frac{3\sqrt{3}(1+8b_1-2{\rm Tr}\Omega^2)}{\sqrt{4{\rm Tr}\Omega^2-1}}\right)^{1/2},\\
\lambda^\pm(+)&=&\frac{1}{4}+\frac{1}{4\sqrt{3}}\sqrt{4{\rm Tr}\Omega^2-1}\pm\frac{1}{2\sqrt{6}}\left(4{\rm Tr}\Omega^2-1-\frac{3\sqrt{3}(1+8b_1-2{\rm Tr}\Omega^2)}{\sqrt{4{\rm Tr}\Omega^2-1}}\right)^{1/2}.
\end{eqnarray}

For both $c_1=0$ and $c_2=0$, the eigenvalues of the $4\times 4$ Hermitian and trace-one matrix are either  
\begin{eqnarray}
\lambda_{1,2,3}&=&\frac{1}{4}-\frac{1}{4\sqrt{3}}\sqrt{4{\rm Tr}\Omega^2-1},\\
\lambda_4&=&\frac{1}{4}+\frac{\sqrt{3}}{4}\sqrt{4{\rm Tr}\Omega^2-1}.
\end{eqnarray}
if $b_0=[3-6\Tr\Omega^2-6(\Tr\Omega^2)^2+\sqrt{3}(4\Tr\Omega^2-1)^{3/2}]/288,\; b_1=[18\Tr\Omega^2-9-\sqrt{3}(4\Tr\Omega^2-1)^{3/2}]/72$, or
\begin{eqnarray}
\lambda_{1,2,3}&=&\frac{1}{4}+\frac{1}{4\sqrt{3}}\sqrt{4{\rm Tr}\Omega^2-1},\\
\lambda_4&=&\frac{1}{4}-\frac{\sqrt{3}}{4}\sqrt{4{\rm Tr}\Omega^2-1}.
\end{eqnarray}
if $b_0=[3-6\Tr\Omega^2-6(\Tr\Omega^2)^2-\sqrt{3}(4\Tr\Omega^2-1)^{3/2}]/288,\;  b_1=[18\Tr\Omega^2-9+\sqrt{3}(4\Tr\Omega^2-1)^{3/2}]/72$.
   
Just is well known, Peres' separability condition tells us, all the eigenvalues of the partial transpose of DM ought to be non-negative \cite{Peres}. Thus, taking the minimum eigenvalue in Theorem one and setting it non-negative, we have

\medskip
{\noindent}{\it Theorem Two}\ The separability condition of DM $\rho$  of two qubits in an arbitrary state is
\begin{eqnarray}
1&\geq&\frac{1}{\sqrt{3}}(4{\rm Tr}\rho^2-1+8c_1^{\rm P}\cos\phi^{\rm P})^{1/2}+\frac{2}{\sqrt{6}}\left[4{\rm Tr}\rho^2-1\right.\nonumber\\
& &\left.-4c_1^{\rm P}\cos\phi^{\rm P}+\frac{3\sqrt{3}(1+8b_1^{\rm P}-2{\rm Tr}\rho^2)}{\sqrt{4{\rm Tr}\rho^2-1+8c_1^{\rm P}\cos\phi^{\rm P}}}\right]^{1/2},
\end{eqnarray}
where
\begin{eqnarray}
b_0^{\rm P}&=&b_0-\frac{1}{32}[(({\rm Tr}A)^2-{\rm Tr}A^2)\bm{\xi}_A \cdot\bm{\xi}_B+2\bm{\xi}_B^T A^2\bm{\xi}_A\nonumber\\ 
& &-2{\rm Tr}A\;\bm{\xi}_B^T A\bm{\xi}_A)]+\frac{1}{16}(\bm{a}_1\times\bm{a}_2)\cdot\bm{a}_3,\label{PTB0}\\
b_1^{\rm P}&=&b_1-\frac{1}{4}(\bm{a}_1\times\bm{a}_2)\cdot\bm{a}_3,\label{PTB1}\\
b_2^{\rm P}&=&b_2,\\
c_1^{\rm P}&=&\sqrt{12b_0^{\rm P}+3b_1^{\rm P}+b_2^{\rm P}{}^2},\label{PTB2}\\
c_2^{\rm P}&=&27b_1^{\rm P}{}^2+b_0^{\rm P}(27-72b_2^{\rm P})+9b_1^{\rm P}b_2^{\rm P}+2b_2^{\rm P}{}^3,\\
\cos\phi^{\rm P}&=&\frac{c_1^{\rm P}}{2^{2/3}(c_2^{\rm P}+\sqrt{c_2^{\rm P}{}^2-4c_1^{\rm P}{}^6})^{1/3}}+\frac{(c_2^{\rm P}+\sqrt{c_2^{\rm P}{}^2-4c_1^{\rm P}{}^6})^{1/3}}{2\times 2^{1/3}c_1^{\rm P}}.
\end{eqnarray}
And $c_1^{\rm P}\neq 0,\;c_2^{\rm P}\neq 0$. 
If only $c_2^{\rm P}=0$, the separability condition becomes
\begin{equation}
1\geq\frac{1}{4\sqrt{3}}\sqrt{4{\rm Tr}\rho^2-1}+\frac{1}{2\sqrt{6}}\left(4{\rm Tr}\rho^2-1+\frac{3\sqrt{3}(1+8b_1^{\rm P}-2{\rm Tr}\rho^2)}{\sqrt{4{\rm Tr}\rho^2-1}}\right)^{1/2}.
\end{equation}
If both $c_1^{\rm P}=0$ and $c_2^{\rm P}=0$, then in case one the DM is always  separable and in case two the separability condition is
\begin{equation}
\Tr\rho^2\leq \frac{1}{3}
\end{equation}
In the above, we have used the fact that the trace of the square of the partial transpose matrix of DM is equal to the trace of the square of DM.   

Obviously, the pure state is the simplest case. In fact, we can prove the following theorem:

\medskip
\noindent {\it Theorem Three}\ The eigenvalues of the partial transpose of DM of two qubits in a pure state $\ket{\phi}=a\ket{00}+b\ket{01}+c\ket{10}+d\ket{11}$ is
\begin{equation}
\mp |ad-bc|,\;\frac{1}{2}(1\mp\sqrt{1-4|ad-bc|^2}),
\end{equation}
and then the separability condition is just
\begin{equation}
ad-bc=0.
\end{equation}
It is consistent with my paper \cite{My0}. 

Because Peres' separability condition is necessary and sufficient one for two qubits, the Theorem two and Theorem three, as the obvious and general expression of Peres' condition, are necessary and sufficient one either.  

If there are some vanishing eigenvalues for a $4\times 4$ Hermitian and trace-one matrix, the conclusions can be simplified. The following theorems will show this judgment. Their proofs can be given by solving the corresponding characteristic equations.

\medskip
\noindent{\it Theorem Four}\ If at least there is one vanishing eigenvalue for a $4\times 4$ Hermitian and trace-one matrix, its eigenvalues are     
\begin{eqnarray}
\lambda_1&=&\frac{1}{3}(1+\sqrt{6{\rm Tr}\Omega^2-2}\; \cos\phi),\\
\lambda_2&=&\frac{1}{3}[1-\sqrt{6{\rm Tr}\Omega^2-2}\; \cos(\phi-\pi/3)],\\
\lambda_3&=&\frac{1}{3}[1-\sqrt{6{\rm Tr}\Omega^2-2}\; \cos(\phi+\pi/3)],
\end{eqnarray}
where
\begin{eqnarray}
\cos\phi&=&
\frac{\sqrt{1-3b_2}}{2^{2/3}(d+\sqrt{d^2-4(1-3b_2)^3})^{1/3}}\\ \nonumber
& &+\frac{(d+\sqrt{d^2-4(1-3b_2)^3})^{1/3}}{2\times 2^{1/3}\sqrt{1-3b_2}}, \label{3COS}\\
d&=&2-27b_1-9b_2.
\end{eqnarray}
Here we have assumed $3{\rm Tr}\Omega^2-1\neq 0$ and $d=(3\lambda_1-1)(3\lambda_2-1)(3\lambda_3-1)\neq 0$. If $d< 0$, then $\pi/6<\phi\leq \pi/3$, and if $d>0$, then $0\leq\phi<\pi/6$. Because that $d^2-4(1-3b_2)^3=-27(\lambda_1-\lambda_2)^2(\lambda_2-\lambda_3)^2(\lambda_3-\lambda_1)^2$, we have $4(1-3b_2)^3\geq d^2 \geq 0$, and $1-3b_2=(3{\rm Tr}\Omega^2-1)/2\geq 0$. If  ${\rm Tr}\Omega^2=1/3$, we have that $d$ has to be zero. Thus, $b_1=-1/27$ and $b_2=1/3$. This implies that all the eigenvalues are equal to $1/3$. In particular,  only $d=0$, the eigenvalues becomes
\begin{eqnarray}
\lambda_1&=&\frac{1}{3},\\
\lambda_{2,3}&=&\frac{1}{3}\left(1\pm\sqrt{\frac{3}{2}}\sqrt{3{\rm Tr}\Omega^2-1}\right).
\end{eqnarray}

\medskip
\noindent{\it Theorem Five}\ If at least there is one vanishing eigenvalue for a $4\times 4$ Hermitian and trace-one matrix, then the positive definite condition of eigenvalues is
\begin{equation}
\sqrt{6{\rm Tr}\Omega^2-2}\; \cos(\phi^{\rm P}-\pi/3)\leq 1.
\end{equation}
If only $d=0$, the positive definite condition becomes
\begin{equation}
{\rm Tr}\Omega^2\leq\frac{5}{9}.
\end{equation} 

\medskip
\noindent{\it Theorem Six}\ If at least there are two vanishing eigenvalues for a $4\times 4$ Hermitian and trace-one matrix, then the other eigenvalues are
\begin{equation}
\lambda_\pm=\frac{1}{2}(1\pm\sqrt{2{\rm Tr}\Omega^2-1}).
\end{equation}
The positive definite condition of the eigenvalues
\begin{equation}
{\rm Tr}\Omega^2\leq 1.
\end{equation}

Thus, from the Peres' separability condition, it is easy to prove the following theorem:

\medskip
\noindent{\it Theorem Seven}\ If the partial transpose of DM of two qubits has at least two vanishing eigenvalues, this density matrix is separable. If the partial transpose of DM of two qubits has only one vanishing eigenvalue, the separability condition is obtained by using of Theorem Five to it and setting the minimum eigenvalue is positive.

Now, let's we discuss some applications of our theorems. Just is well known that  many measures of entanglement are related with the quantum entropies which are defined by density matrix, for example, the entanglement of formation \cite{Bennett} and the relative entropy of entanglement \cite{Vedral}. To compute quantum entropy,    we often need to find the eigenvalues of density matrix. Even, according to Wootters \cite{Wootters}, the measure of entanglement of two qubits is directly determined by the eigenvalues of $4\times 4$ hermitian matrix. Therefore, in terms of our Theorems about the eigenvalues, we can easily calculate the entanglement of formation of an arbitrary state of two qubits. As to the relative entropy of entanglement or its improving \cite{My1}, we have to calculate von Neumann entropy which is just defined directly by the eigenvalues.

Furthermore, we can find a relation between the upper bound of entanglement and the eigenvalues.

\medskip
\noindent{\it Theorem Eight}\ If all the eigenvalues of DM of two qubits are not zero, the possibly maximum value of the entanglement of formation is not larger than
\begin{eqnarray}
& &\frac{1}{\sqrt{3}}(4{\rm Tr}\rho^2-1+8c_1^{\rm P}\cos\phi^{\rm P})^{1/2}+\frac{2}{\sqrt{6}}\left[4{\rm Tr}\rho^2-1\right.\nonumber\\
& &\left.-4c_1^{\rm P}\cos\phi^{\rm P}+\frac{3\sqrt{3}(1+8b_1^{\rm P}-2{\rm Tr}\rho^2)}{\sqrt{4{\rm Tr}\rho^2-1+8c_1^{\rm P}\cos\phi^{\rm P}}}\right]^{1/2}.
\end{eqnarray}

This is because DM of two qubits can be written as
\begin{equation}
\rho=\lambda_{\min} I+(1-4\lambda_{\min})\rho^\prime.
\end{equation}
Note that we can not put a number larger than $\lambda_{\min}$ in front of the identity matrix $I$, because we have to keep $\rho^\prime$ to be positive definite.

Furthermore, let's consider a model of transfer of entanglement \cite{My2}. This can be expressed as a following story. Alice and Bob are friends. One day, Alice and Bob sat together at the lounge in a party. On the left side of Alice is Charlie and on the right side of Bob is David. Alice and Charlie, Bob and David respectively exchanged their seats. This leads to Alice and Bob's entanglement decreases. In language of quantum information, Alice and Bob shares an entangled state initially. Without loss of generality, suppose it is in Bell state $(\ket{00}+\ket{11})/\sqrt{2}$, Charlie is in $\ket{c}$ and David is in $\ket{d}$. That is that four of them is in a total state $\ket{c}\otimes(\ket{00}+\ket{11})\otimes\ket{d}/\sqrt{2}$. Now, introduce the swapping interaction\cite{Loss} respectively between Alice and Charlie, and between Bob and David:
\begin{equation}
S=\left(\begin{array}{cccc}
1&0&0&0\\
0&0&1&0\\
0&1&0&0\\
0&0&0&1
\end{array}\right),\qquad S\ket{ab}=\ket{ba}\quad (a,b=0,1).
\end{equation}
It is easy to see that
\begin{equation}
S\otimes S\left[\frac{1}{\sqrt{2}}\ket{c}\otimes(\ket{00}+\ket{11})\otimes\ket{d}\right]=\frac{1}{\sqrt{2}}(\ket{0}\otimes\ket{cd}\otimes\ket{0}+\ket{1}\otimes\ket{cd}\otimes\ket{1}).
\end{equation}
Thus, the entanglement between the second qubit and the third qubit is transfered to the entanglement of the first qubit and the fourth qubit. In general, the swapping process suffers the affection of noisy. We suppose, after transfer of entanglement,  that DM becomes   
\begin{equation}
\rho^\prime=(1-\epsilon)\rho+\epsilon\frac{1}{4}I. 
\end{equation}
where $\epsilon$ represents the strength of the noise and $\rho$ is, in form, the same as DM before transfer of entanglement. Obviously, this model can be extended to a qubit chain:
\begin{equation}
\cdots\overbrace{\underbrace{\bullet\leftrightarrow\cdots\leftrightarrow\bullet\leftrightarrow}_n\overbrace{\bullet\leftrightarrow\underbrace{\bullet\qquad\bullet}_{\rho_0}\leftrightarrow\bullet}^{\rho_1}\underbrace{\leftrightarrow\bullet\leftrightarrow\cdots\leftrightarrow\bullet\leftrightarrow\bullet}_n}^{\rho_n}\cdots
\end{equation}
Through the swapping interaction, the entanglement can be transfered along with the chain one node by one node and forward to the opposite directions. At the beginning, denote DM for a pair of given adjacent nodes of a qubit chain as $\rho_0$. After the first swapping, the DM of a pair of qubits respectively on the nodes of the left side and right side of the given pair of qubits is written as $\rho_1$. Since the affection of noise, after the $n$-th swapping in turn along with two directions, the DM of a pair of qubits respectively on the $n$-th nodes of the left side and right side of the original two adjacent qubits becomes
\begin{equation}
\rho_n=(1-\epsilon)\rho_{n-1}+\epsilon\frac{1}{4}I. 
\end{equation}
This equation has not taken into account the fact that $\rho_n$ and $\rho_{n-1}$ are related with the different pairs of qubits, and let it is valid only in mathematics. We also assume that at the beginning a pair of qubits in the adjacent nodes is in a pure state. Thus we would like to know what is $n$'s value if $\rho_n$ is separable. That is to calculate the transfer distance $n$ of entanglement for such a chain of qubits through a noisy channel. 

According to our theorem, the minimum eigenvalue of the partial transpose of $\rho_n$ is 
\begin{equation}
\lambda_{\min}=\frac{1}{4}[1-((1-\epsilon)^n(1+4|ad-bc|)].
\end{equation}
By using of Peres' criterion for the separable state, we can find that
\begin{equation}
n\leq-\frac{\log(1+4|ad-bc|)}{\log (1-\epsilon)}.
\end{equation}
In particular, when $\rho_0$ is a density matrix in the maximum entangled state, we obtain
\begin{equation}
n\leq\displaystyle -\frac{\log 3}{\log (1-\epsilon)}.
\end{equation}
Obviously, when one hopes $n=10$, then it allows the noisy strength $\epsilon$ is not larger than $0.104042$. When the noisy strength $\epsilon$ is larger than $0.42265$, any transfer will lead in disentanglement. If the noisy strength $\epsilon$ only reaches at $0.01$ or $0.1$, the transfer distance $n$ can be 109 or 10. Likewise, we can describe the transfer of entanglement along with one direction. The significances and applications of this model of transfer of entanglement should be imaginable.

In addition, we hope to apply our theorems into seeking the minimum pure state decomposition of DM, this work is in progressing. In a words, we can say, of course, the theorems proposed here are the useful tools to study the entanglement and the related problems. 

\smallskip  
I would like to thank Artur Ekert for his great help and for his hosting my visit to center of quantum computing in Oxford University.

\end{document}